\documentstyle[honnef,pslatex,graphics,graphicx]{article}
\frompage{000} \topage{000}                                              

\newcommand{\ap}{\mathrm{\overline{p}}}
\newcommand{\ad}{\mathrm{\overline{d}}}

\newcommand{\ahe}{\mathrm{\overline{^3He}}}

\def\rightmark{Antimatter and Matter Production in Heavy Ion Collisions at CERN}
\def\leftmark{K. Pretzl et al.}
 
\title{Antimatter and Matter Production in Heavy Ion Collisions at CERN
(The NEWMASS Experiment NA52)}

\authors{
{
 {\em K. Pretzl for the NA52 collaboration:}\\[2mm]
 G.~Ambrosini$^1$, 
 R.~Arsenescu$^1$, 
 C.~Baglin$^2$,
 H.P.~Beck$^1$,
 K.~Borer$^1$,
 A.~Bussi\`ere$^2$,
 K.~Elsener$^3$,
 Ph.~Gorodetzky$^5$,
 J.P.~Guillaud$^2$,
 P.~Hess$^1$,
 S.~Kabana$^1$,
 R.~Klingenberg$^1$,
 G.~Lehmann$^1$,
 T.~Lind\'en$^4$,
 K.D.~Lohmann$^3$,
 R.~Mommsen$^1$,
 U.~Moser$^1$,
 K.~Pretzl$^1$,
 J.~Schacher$^1$,
 R.~Spiwoks$^1$,
 F.~Stoffel$^1$,
 J.~Tuominiemi$^4$,
 M.~Weber$^1$
}\\[2.812mm]
{\normalsize
\hspace*{-8pt}$^1$ Laboratory for High Energy Physics, University of Bern, 
Sidlerstrasse~5, CH-3012 Bern, Switzerland\\ [0.2ex]
\hspace*{-8pt}$^2$ CNRS-IN2P3, LAPP Annecy, F-74941 Annecy-le-Vieux, Fr
ance\\ [0.2ex]
\hspace*{-8pt}$^3$ CERN, SL Division, CH-1211 Geneva 23, Switzerland\\ [0.2ex]
\hspace*{-8pt}$^4$ Dept. of Physics, University of Helsinki, PO Box 9,
FIN-00014 Helsinki, Finland\\ [0.2ex]
\hspace*{-8pt}$^5$ PCC - College de France, 11 Place Marcellin Berthelod
 , 75005 Paris, France\\ [0.2ex]
}}

\abstract{
Besides the dedicated search for strangelets NA52 measures 
light (anti)particle and (anti)nuclei production over a wide range of 
rapidity.
Compared to previous runs the statistics has been increased in the 1998 run by more than one order of magnitude for negatively charged objects at different spectrometer 
rigidities. 
Together with previous data taking at a rigidity of $-20$ GeV/$c$
 we obtained $10^6 \; \ap$, $10^3 \;\ad$ and two $\ahe$ without 
centrality requirements.
We measured nuclei and antinuclei (p,d,$\overline{p}$, $\overline{d}$)
near midrapidity  covering
an impact parameter range  of b$\sim$2-12 fm. 
Our results strongly indicate that nuclei and antinuclei
are mainly produced via the coalescence mechanism.
However the centrality dependence of the 
antibaryon to baryon ratios show that antibaryons
are diminished due to annihilation and breakup reactions in the 
hadron  dense environment.
The volume of the particle source extracted from coalescence
models agrees with results from pion interferometry
for an expanding source.
The chemical and thermal freeze-out of nuclei and antinuclei 
appear to coincide with each 
other and with the thermal freeze-out of hadrons.
}

\begin{document}

\maketitle
\vspace*{24pt}

\section{Introduction}
NA52 is a fixed target experiment at the CERN SPS looking at 158 $A$ GeV/$c$ Pb-Pb collisions.
We identify single particles near $p_{\perp}=0$.
Our apparatus is sensitive to all objects reaching the trigger counter located $0.9 \mathrm{ \mu s} \cdot c$ behind the target. 
Besides the dedicated search for strangelets 
\cite{str,s96,qm96,reiner_moriond,sqm2000}
 we measure (anti)particle and (anti)nuclei over a wide range of rapidity 
\cite{sqm2000,qm97,padova1,padova2,part,antinucl,michele_qm99,my_sqm2000}.
We have recorded $10^6 \; \ap$, $10^3 \;\ad$ and two $\ahe$ at
a spectrometer rigidity of $-20$ GeV/$c$.
No antitriton has been observed.
This finds its explanation in the factor of four 
smaller acceptance for a singly charged particle like the triton
as compared to a doubly charged particle like the
$\ahe$, with approximately
the same production cross section.
\\

\noindent
In the present paper we focus on features of baryon, antibaryon as well
as nuclei
and antinuclei production in Pb+Pb collisions at 158 A GeV.
A general description of the experimental setup can be found 
in \cite{sqm2000,michele_qm99,my_sqm2000,chef}.
The $m_T$ and  rapidity dependence of baryons and antibaryons
and their comparison to coalescence and thermodynamical 
model predictions and to p+Be collisions
at 220 GeV, are shown here for the first time.
Antimatter production in heavy ion collisions, may 
give important experimental information on the event of the QCD phase transition
of deconfined quarks and gluons to confined hadrons \cite{prediction}.
The discovery of the latter is an outstanding goal of the heavy ion
physics experiments   \cite{press}.
\\

\begin{figure}[htb]
\begin{center}
\includegraphics*[width=6cm]{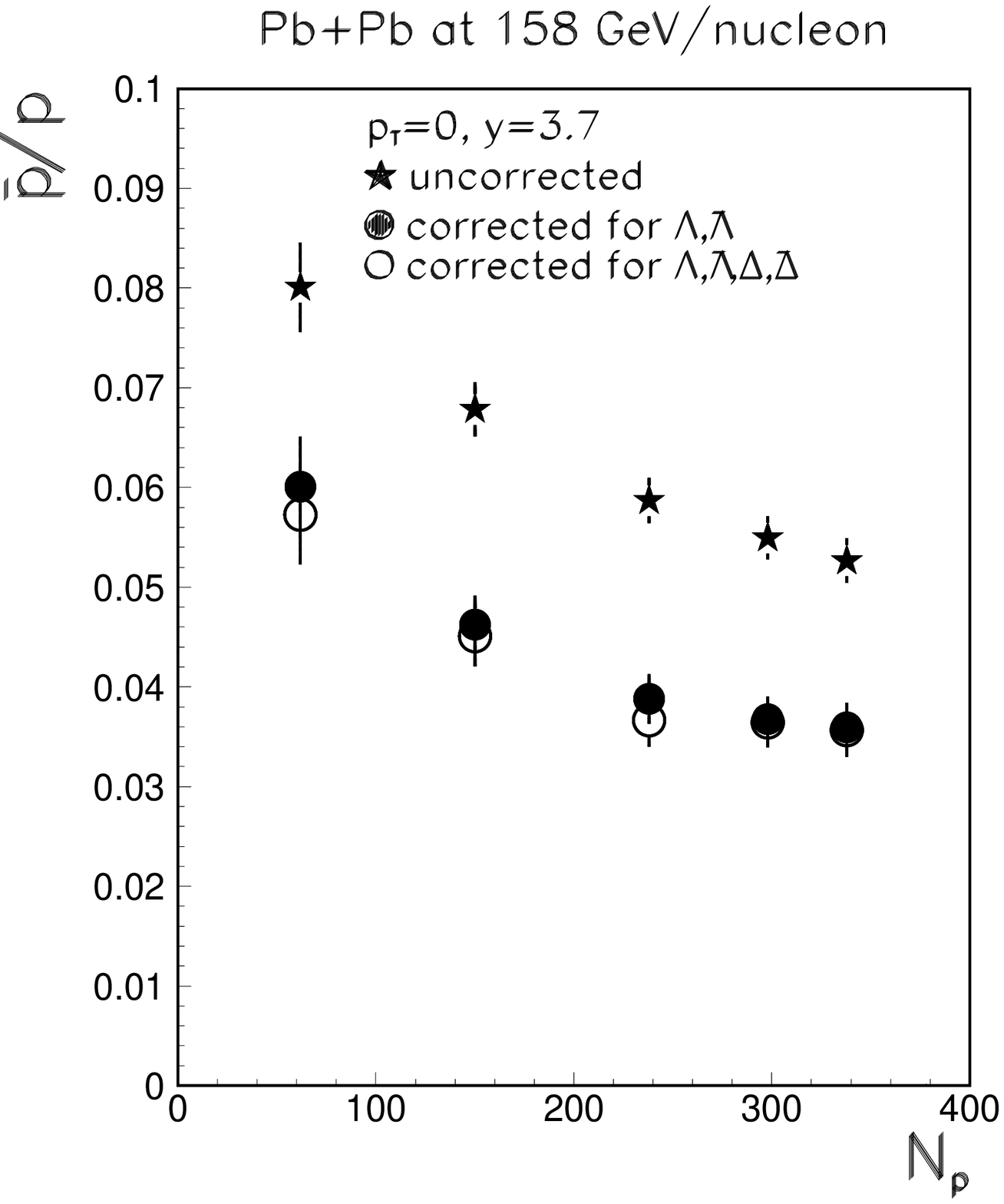}
\includegraphics*[width=6cm]{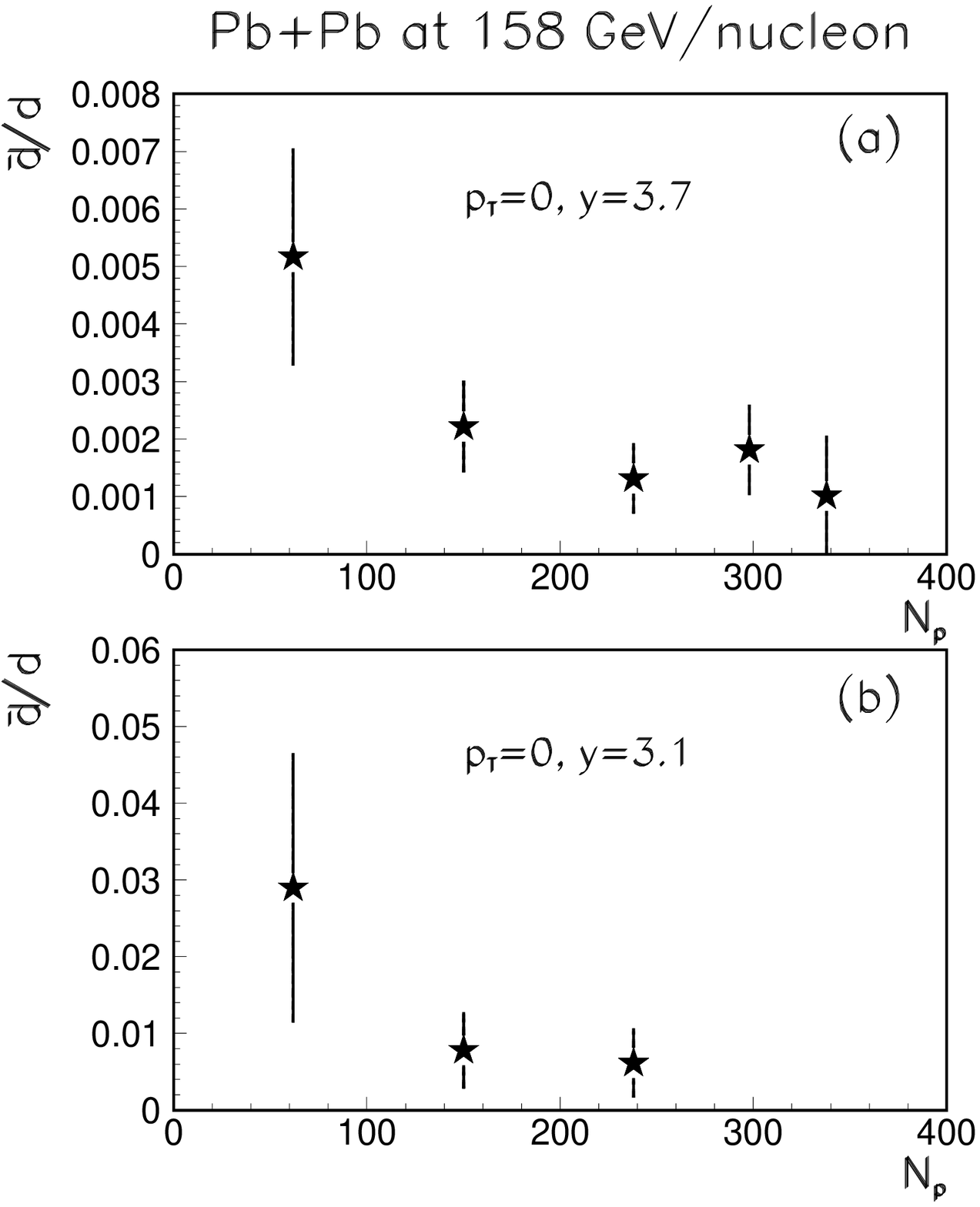}
\end{center}
\caption[]{
Dependence of the $\overline{p}/p$ (left) and the $\overline{d}/d$ (right)
ratios
from the mean number of participant nucleons
 in Pb+Pb collisions at 158 A GeV near zero $p_T$.
The $\Lambda$ and $\Delta$
correction is performed using VENUS 4.12 \protect\cite{venus412}
}
  \label{abtob}
\vspace*{-1cm}
\end{figure}

\begin{figure}[htb]
\begin{center}
\includegraphics*[width=5cm]{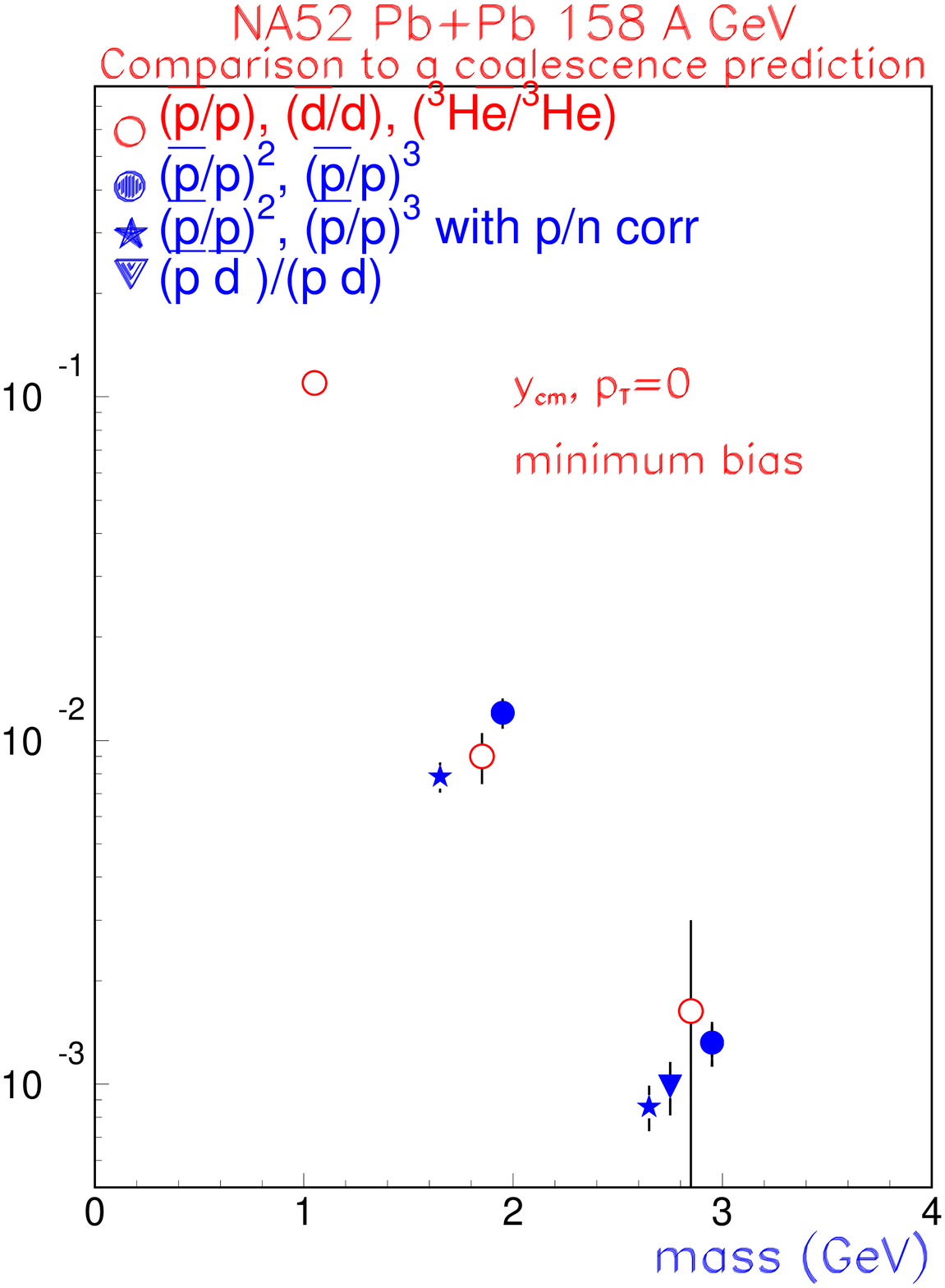}
\includegraphics*[width=5cm]{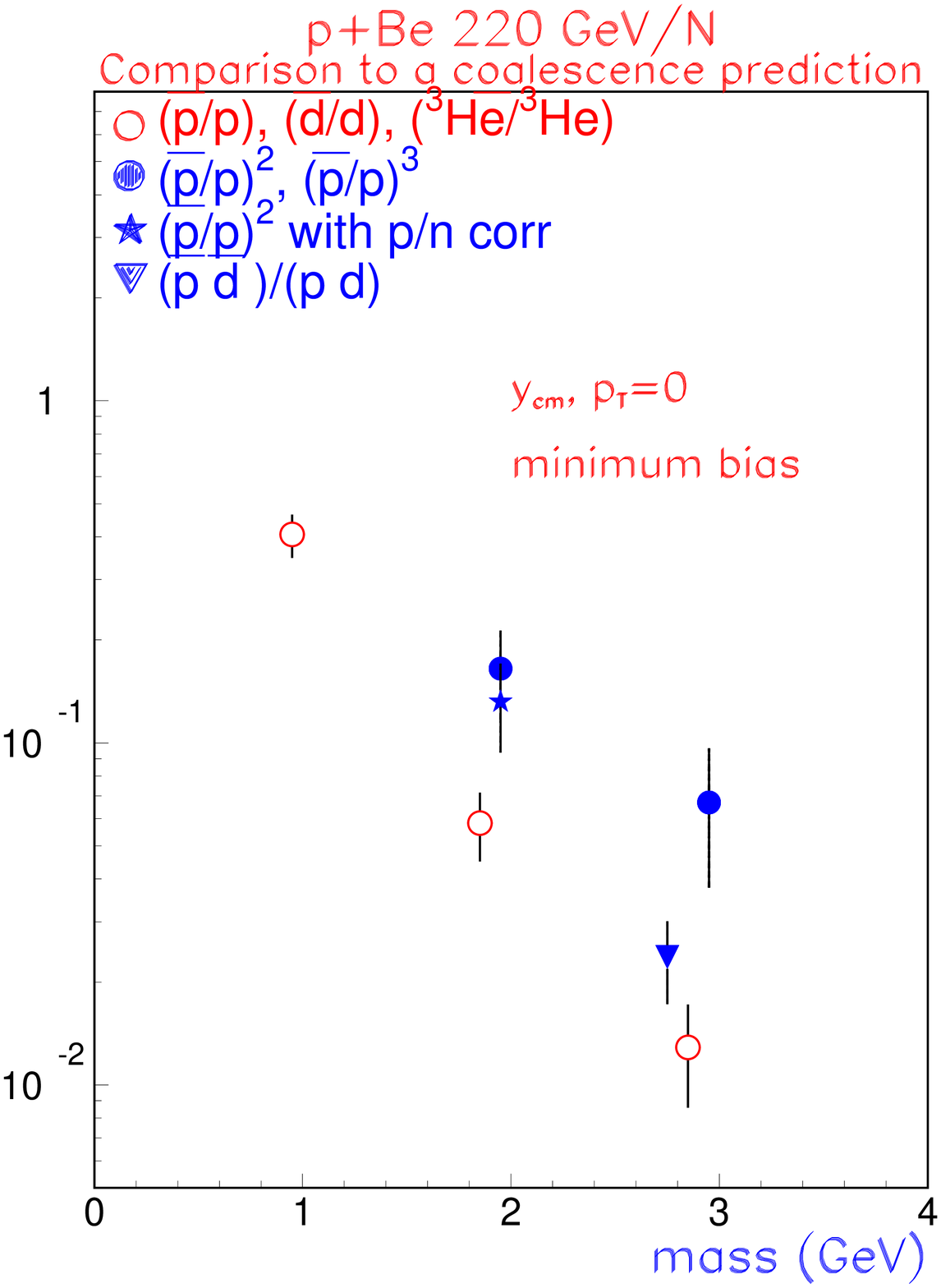}
\end{center}
\caption[]{
Antiparticle to particle ratios near zero transverse momentum and 
at midrapidity, compared to a coalescence model prediction
as a function of particle mass.
Left: data from Pb+Pb collisions at 158 GeV per nucleon (NA52).
Right: data from p+Be collisions at 220 GeV \protect\cite{pbe}.
All data are minimum bias. See text for explanation.
}
\label{coal}
\end{figure}

\begin{figure}[htb]
\vspace*{0.4cm}
\vspace*{0.4cm}
\vspace*{0.4cm}
\vspace*{0.4cm}
\vspace*{-0.4cm}
\vspace*{-0.4cm}
\vspace*{-0.4cm}
\begin{center}
\vspace*{-0.4cm}
\includegraphics*[width=9cm]{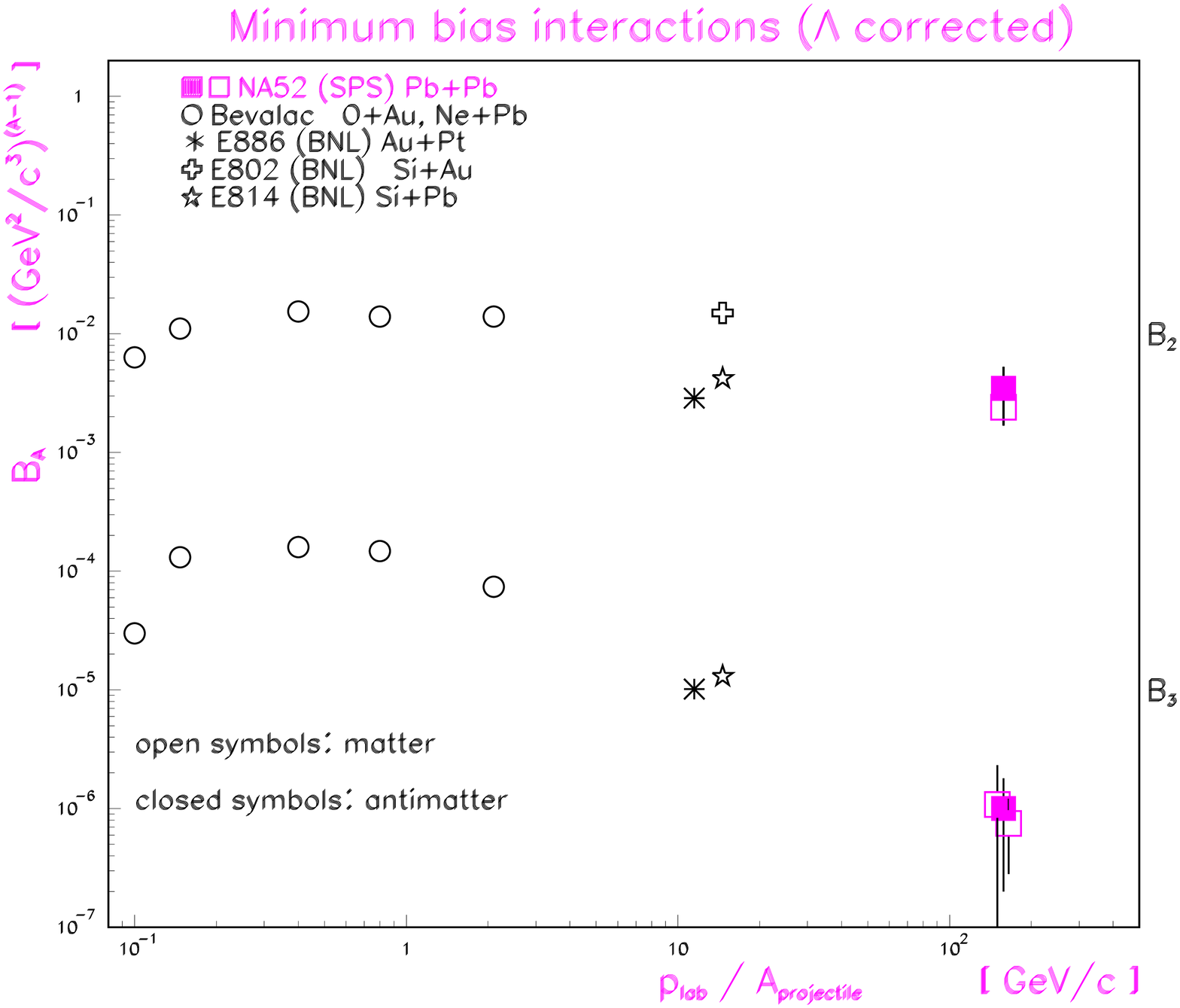}
\vspace*{0.3cm}
\includegraphics*[width=9cm]{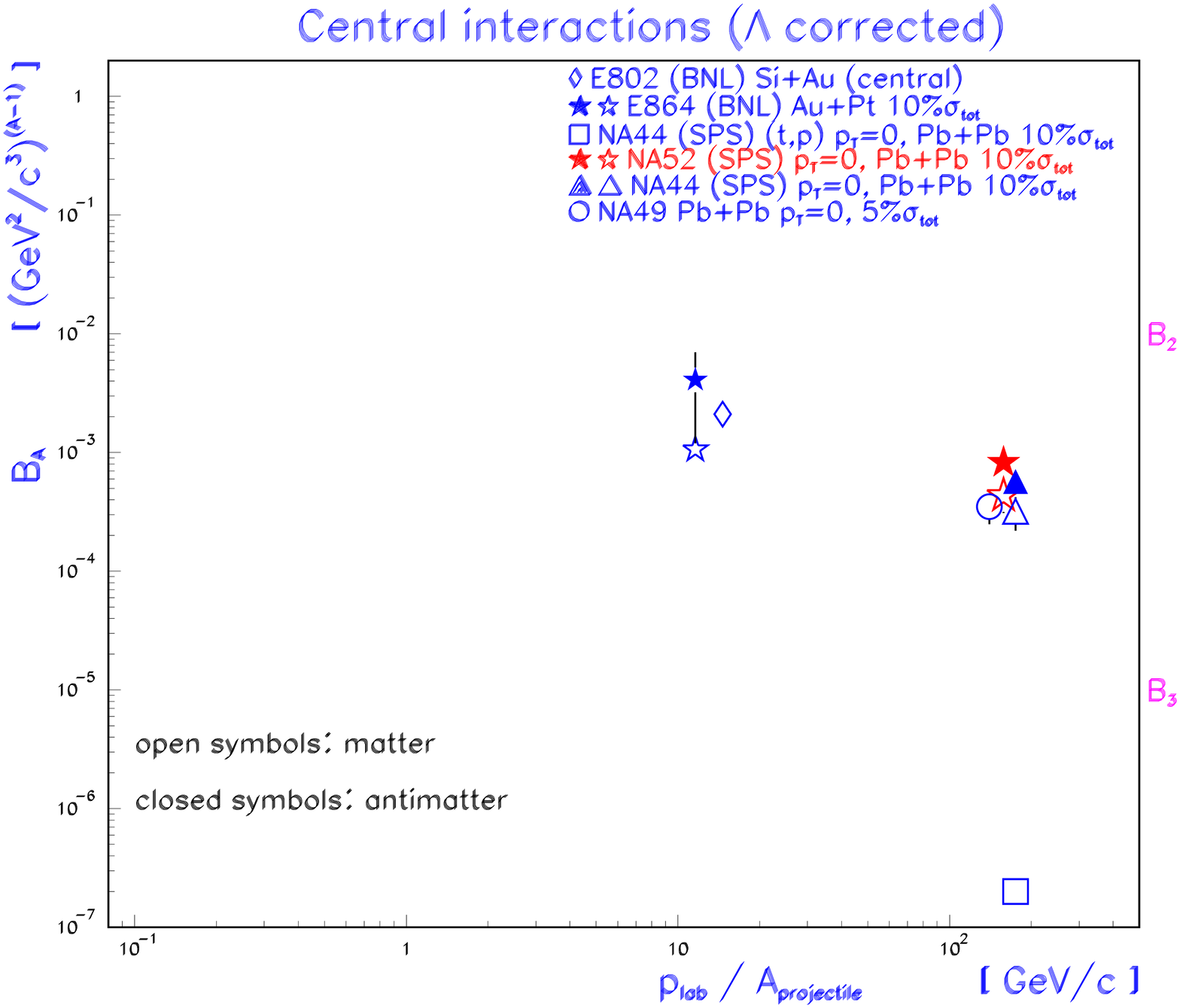}
\end{center}
\vspace*{-1cm}
\caption[]{
 Compilation of coalescence scaling factors at different momentum per nucleon of the incident projectile. 
The above figure shows data taken with minimum bias trigger (all impact
parameters), the one
 below with central trigger (small impact parameters, typically
5-10\% of $\sigma_{tot}$).
The $d/p^2$ and $\overline{d}/ \overline{p}^2$ data ($B_2$) of NA52
 are corrected for $\Lambda$ and $\overline{\Lambda}$ decays.
}
\vspace*{-1cm}
  \label{compil}
\end{figure}

\begin{figure}[htb]
\begin{center}
\includegraphics*[width=5cm]{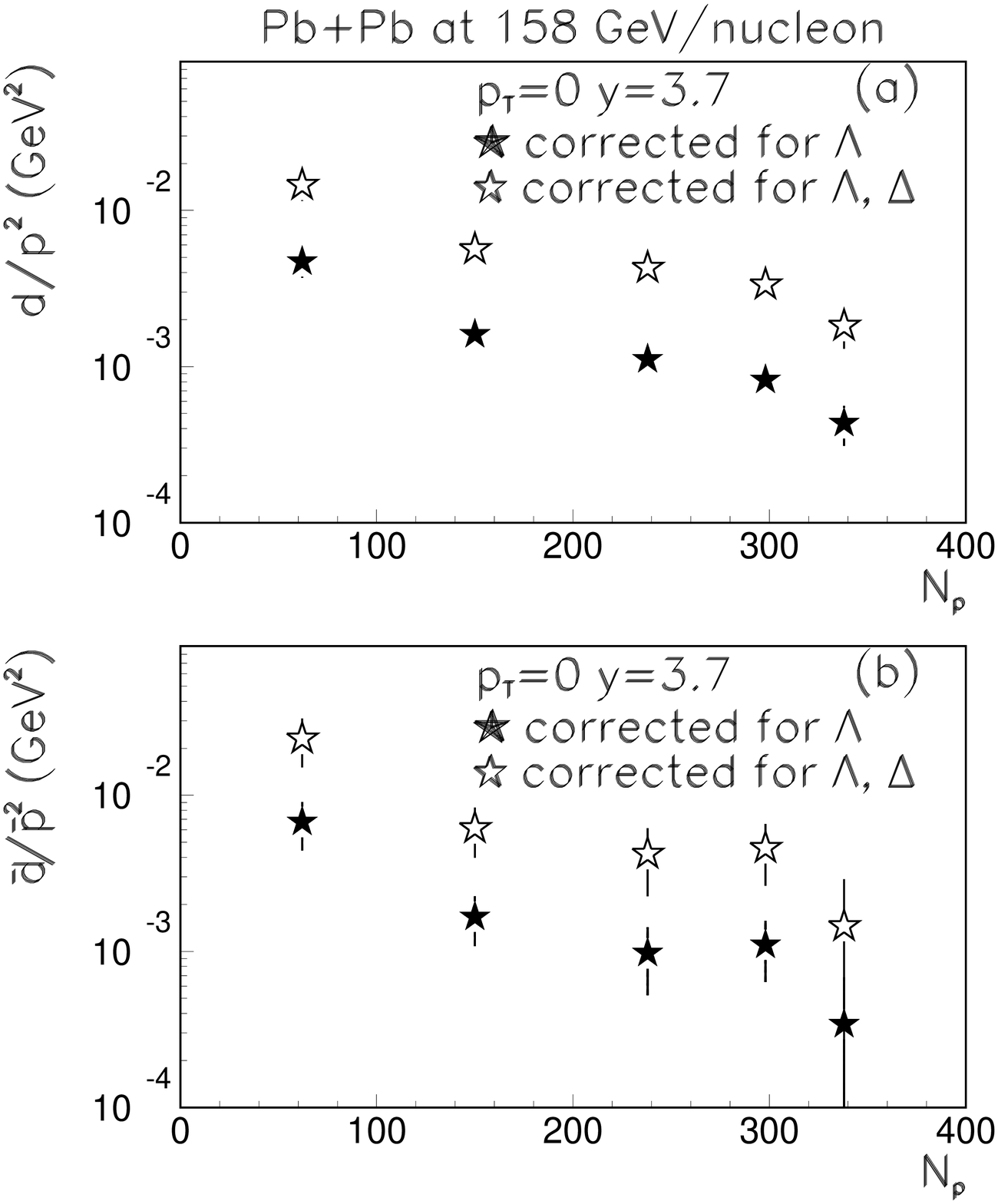}
\hspace*{0.6cm}
\includegraphics*[width=5cm]{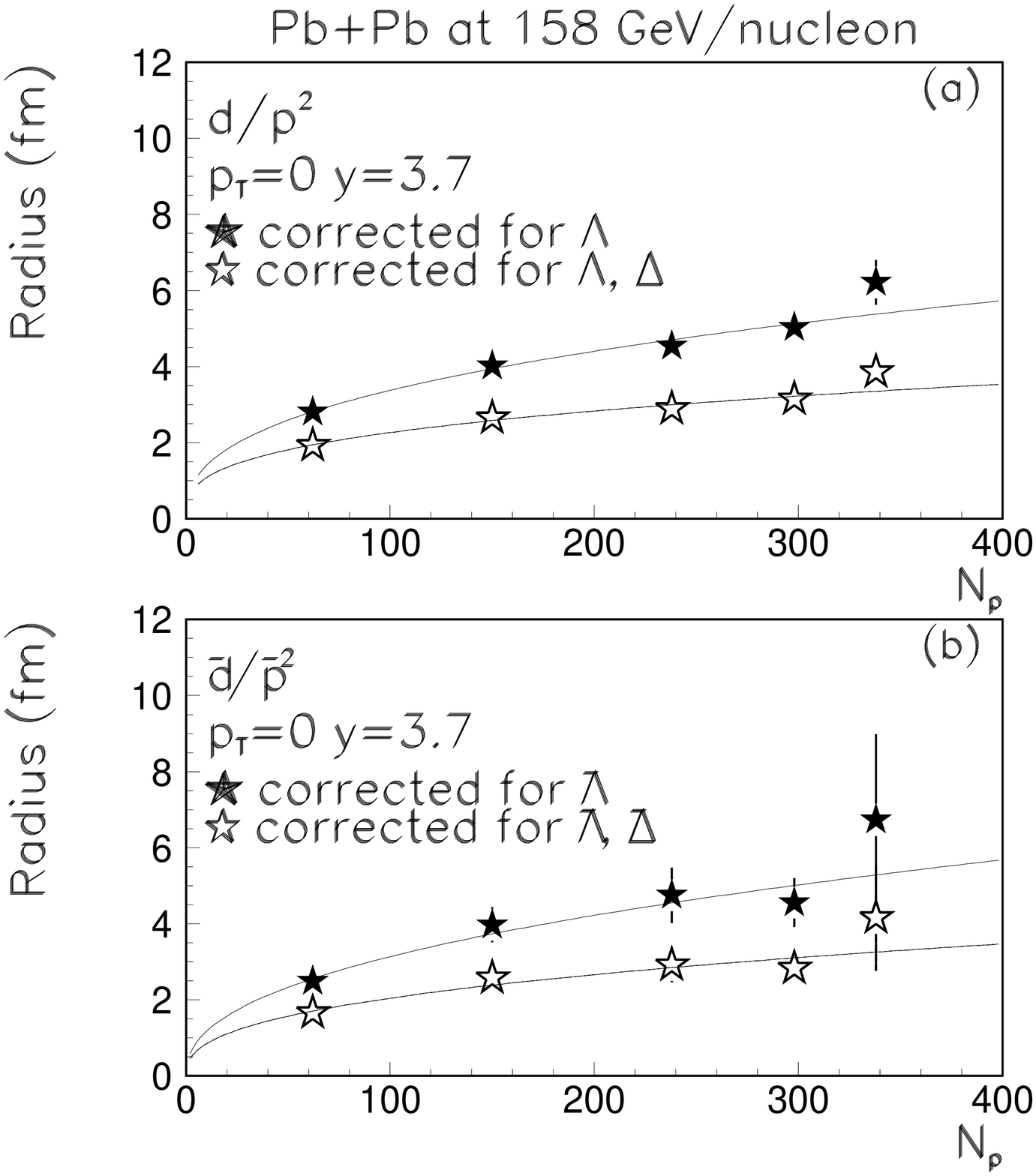}
\end{center}
\caption[]{
Data from Pb+Pb collisions at 158 A GeV.
$d/p^2$  and $\overline{d}/ \overline{p}^2$ yield ratios (left)
and radii extracted from them (right),
 as a function of the number
of participant nucleons $N_p$. 
}
\label{r1}
\vspace*{-1cm}
\end{figure}

\begin{figure}[htb]
\begin{center}
\includegraphics*[width=5cm]{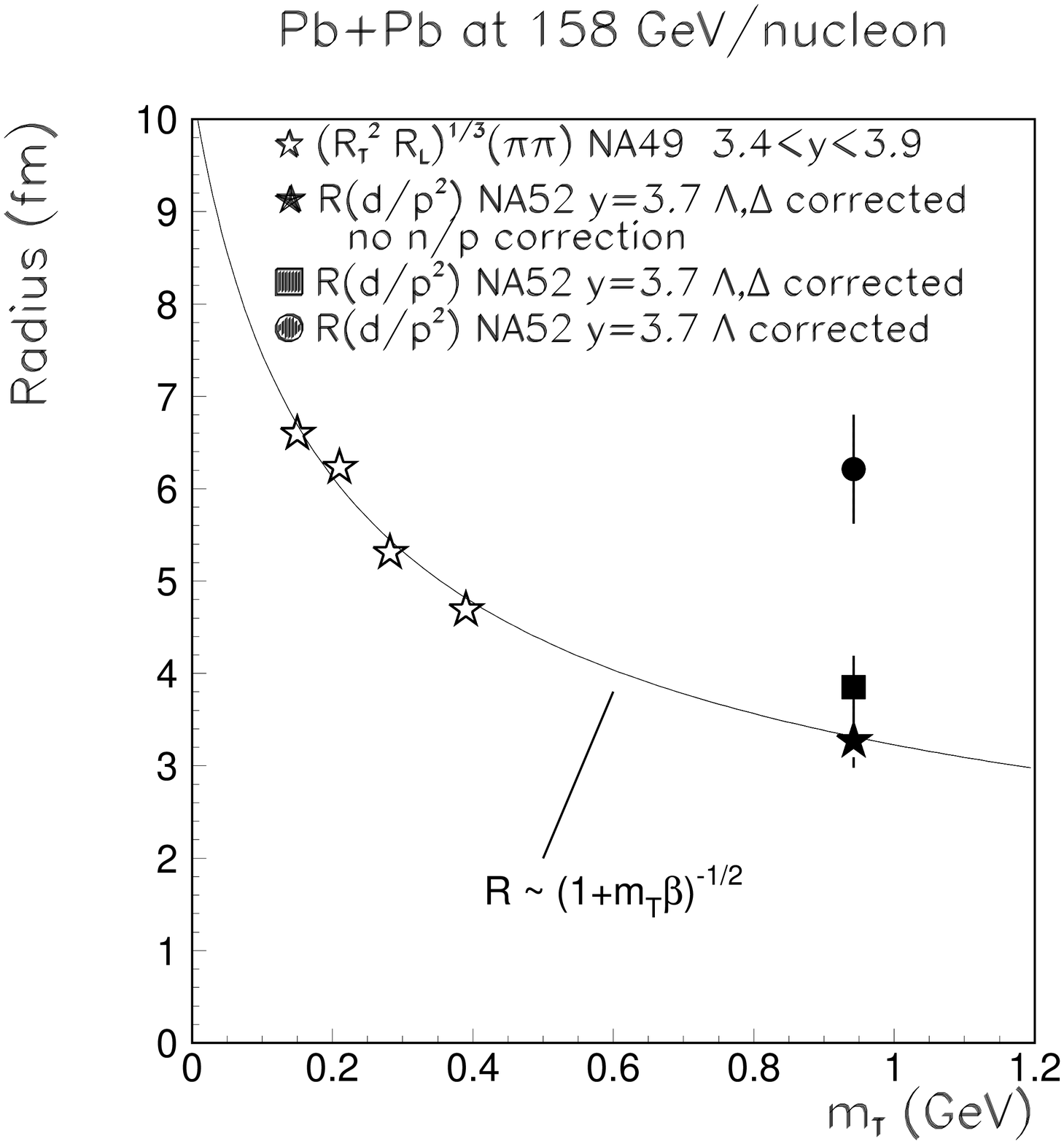}
\hspace*{0.6cm}
\includegraphics*[width=5cm]{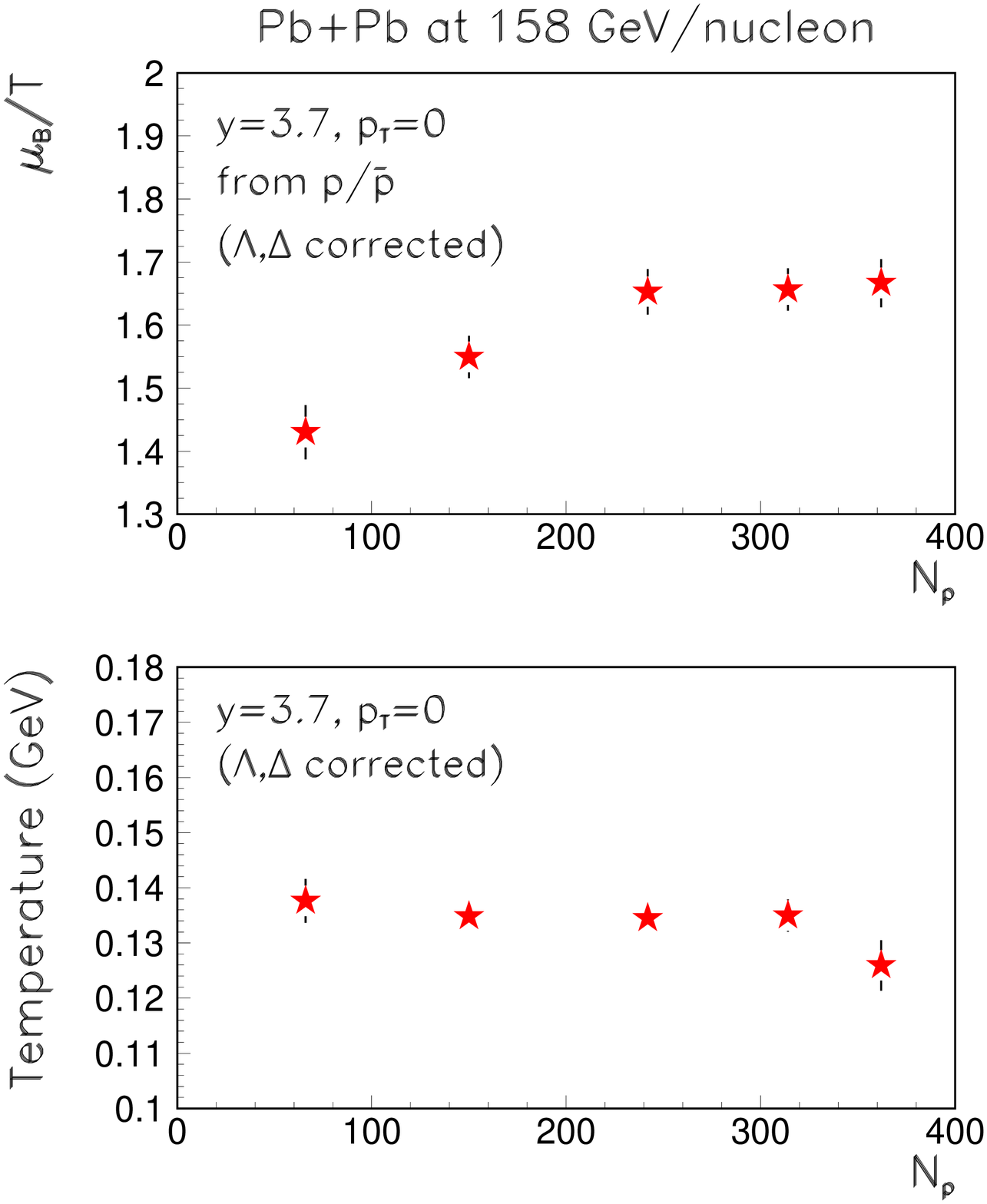}
\end{center}
\caption[]{
Left:
Transverse mass ($m_T=\sqrt{ p_T^2 + m^2}$)
 dependence of the source radii extracted from
$d/p^2$ at y=3.7 (full points, NA52 experiment) and of those
 extracted from $\pi \pi$-correlations
at 3.4$<y<$3.9 (open points, NA49 experiment) in
Pb+Pb collisions at 158 GeV per nucleon.
Right: The baryochemical potential $\mu_B$ over the temperature $T$ 
 at y=3.7 and near zero $p_T$ in Pb+Pb collisions at 158 A GeV,
 as a function of the mean number of participant nucleons.
}
  \label{t_vs_n}
\vspace*{-1cm}
\end{figure}

\begin{figure}[htb]
\begin{center}
\includegraphics*[width=5cm]{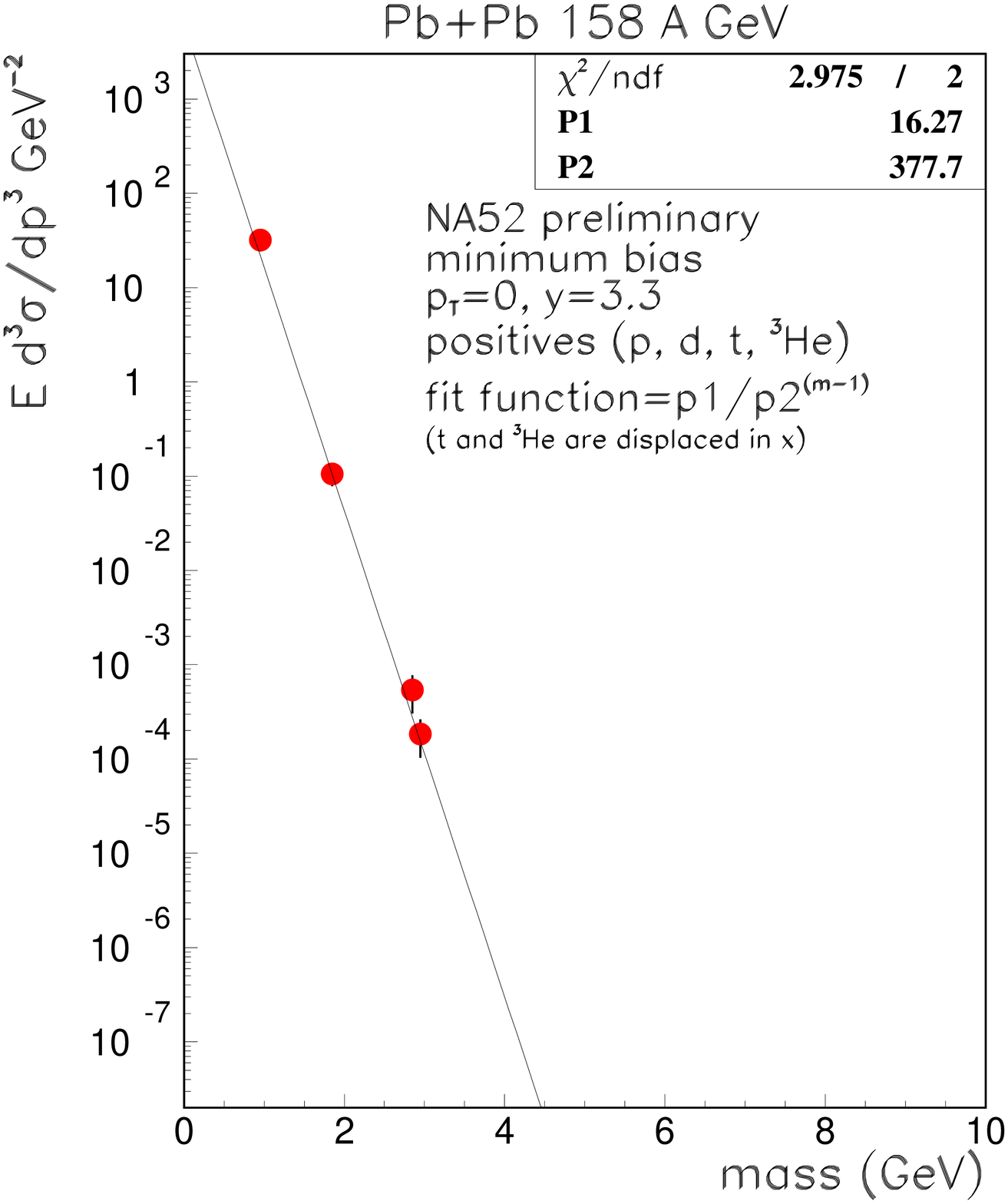}
\includegraphics*[width=5cm]{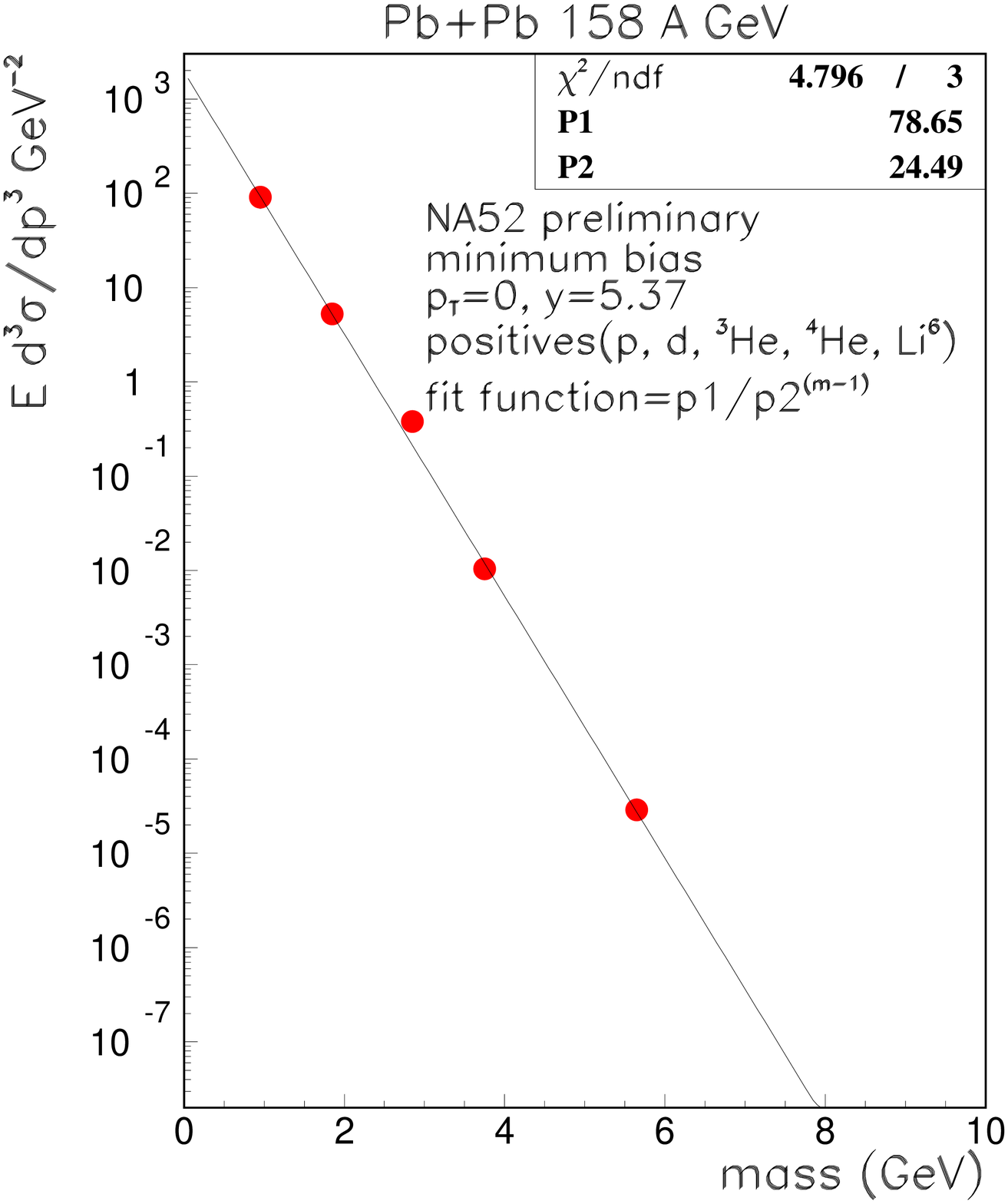}
\end{center}
\caption[]{
Proton and nuclei cross sections as a function of mass in Pb+Pb
collisions at 158 A GeV taken with a minimum bias trigger and
near zero $p_T$.  Left data at y=3.3 and right at y=5.37.
The systematic errors have been quadratically added to the statistical ones.
The resulting  errorbars are of the size of the data points.
The straight lines represent a fit through the data points
using the function $f= p1/p2^{( m-1) }$.
}
\label{nagle1}
\vspace*{-1.cm}
\end{figure}

\begin{figure}[htb]
\begin{center}
\includegraphics*[width=5cm]{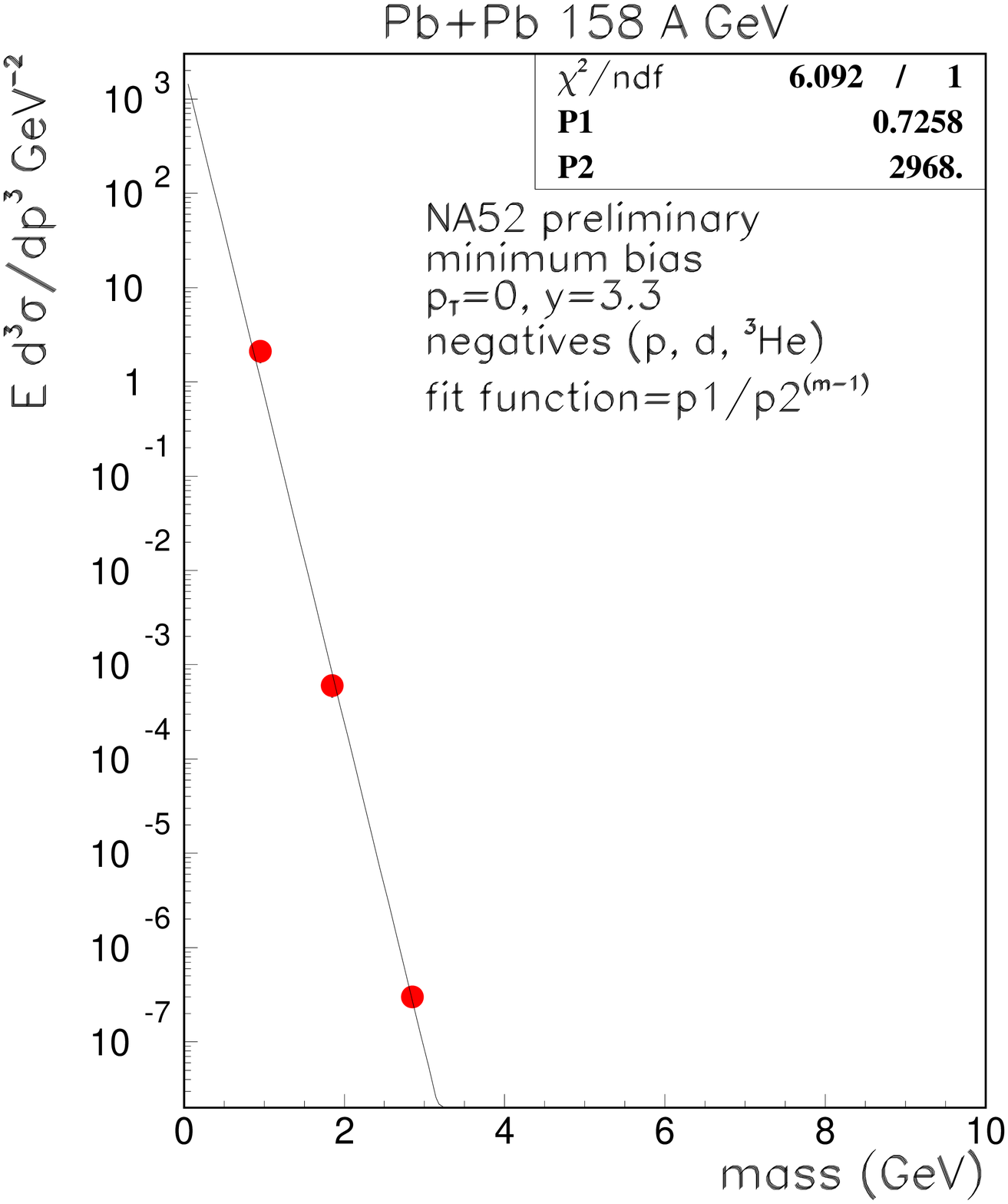}
\end{center}
\caption[]{
Antiproton and antinuclei cross sections as a function of mass in Pb+Pb
collisions at 158 A GeV taken with a minimum bias trigger and
near zero $p_T$ and at y=3.3.
The systematic errors have been quadratically added to the statistical ones.
The resulting  errorbars are of the size of the data points.
The straight lines represent a fit through the data points
using the function $f= p1/p2^{ (m-1) }$.
}
\label{nagle2}
\vspace*{-1.cm}
\end{figure}

\vspace*{1.cm}

\section{Results and discussion}

\noindent
Antibaryon production is expected to be enhanced in a collision
which goes through the QCD phase transition, as compared to one which
does not \cite{prediction}.
However a possible enhancement can  be counterbalanced by the effect
of annihilation of antibaryons in the baryon rich environment during
the course of a collision of heavy nuclei like lead
\cite{annihil}.
In lead lead collisions at $\sqrt{s}$=17 GeV,
baryons are stopped enough to populated significantly
the central rapidity region \cite{padova1}.
\\

\noindent
The centrality dependence of particle, antiparticle and baryon,
antibaryon yields
has previously been investigated by the NA52 experiment
and is published in \cite{padova1,my_sqm2000}.
A comparison of our data to the results of other experiments
can be found in \cite{kaons}.
\\

\noindent
In Fig. \ref{abtob} our  data on $\overline{p}/p$ and
$\overline{d}/d$ ratios
as a function of centrality are shown.
Throughout this paper the shown errorbars correspond to statistical errors
only, unless stated differently.
The systematic error which is mainly due to uncertainties
in the acceptance of the spectrometer is estimated to be 20\%.
The observed decrease of both ratios with increasing
centrality of the collision can be understood as a result
of increasing annihilation of antibaryons
in a baryon rich environment of a central collision.
\\

\noindent
However this effect may also be 
 connected to the low $p_T$ acceptance of our spectrometer.
Preliminary analysis of 
other measurements at full $p_T$ and in a similar centrality range
 show no significant decrease of the $\overline{p}/N$ ratio with 
centrality \cite{sikler}.
It is conceivable that near zero $p_T$ the baryon density is higher
than at larger $p_T$ values,
due to many projectile fragments, which acquire a very small
$p_T$ kick, but populate a large range in rapidity.
These small angle protons can induce a larger annihilation of the antiprotons
which are produced near zero $p_T$.
\\

\noindent
At the moment it is not possible to speak about antibaryon enhancement
in the investigated collisions, because the effect of annihilation
has not yet been quantified.
NA52  took data in 1998 with the goal 
 to measure the effect of antibaryon annihilation
through measurements of anisotropic distributions of antibaryons
in and out of the 'event plane' 
(see \cite{qm97} for first results).
\\

\noindent
Other questions which arise are
how and when are the nuclei and antinuclei produced in the course
of the collision, and what can we learn from their properties.
We investigate first their production mechanism.
Besides direct pair production,
nuclei and antinuclei can also be produced by coalescence
of $p, n, \overline{p}, \overline{n}$ and of other light nuclei and antinuclei.
But also other mechanisms like collective antimatter production
in analogy to spontaneous positron emission and vacuum decay
processes in QED may play a role \cite{new}.
Furthermore, nuclei can  originate from projectile/target
fragmentation which, although they peak at beam/target rapidity and zero
$p_t$,  can populate the whole rapidity region due to stopping.
\\

\noindent
We address the question if coalescence is the dominant production
mechanism by comparing the 
$\overline{d}/d$ and $\overline{^3He}/ ^3He $
ratios with simple coalescence model
 predictions using the data themselves, namely
we examine if the following conditions for the yield ratios hold:
\\ 
($\overline{p}/p$)$^2$  $\sim$ $\overline{d}/d$
\\
($\overline{p}/p$)$^3$ $\sim$ $\overline{^3He}/ ^3He $
\\
$( \overline{p} \overline{d} )/  (pd)$  $\sim$ $\overline{^3He}/ ^3He $.
\\
The results are shown in figure \ref{coal}.
We assumed that the number of $n, \overline{n}$ is equal
to the number of $p, \overline{p}$.
An exception are the 
points noted as 'with p/n correction' where we include a correction
to the number of neutrons using $n = p \cdot (A-Z/Z)$. 
This correction acounting for the number of protons and
neutrons in the target and projectile nuclei
is not necesserily correct since the ratio
$n/p$ may change in the course of the collision.
The coalescence prediction agrees with the  Pb+Pb data at
158 A GeV (figure \ref{coal} (left)), while it does not
describe so well the p+Be data at 220 GeV  (figure \ref{coal} (right)).
Both data sets are minimum bias, at $p_T \sim 0$ and at midrapidity.
Figure \ref{coal} suggests that coalescence is the dominant 
production mechanism for nuclei and antinuclei in Pb+Pb collisions,
 while in p+Be collisions
 nucleon and antinucleon pairs seem to be produced more directly and
are less affected by annihilation processes and breakup.
\\

\clearpage

\begin{figure}[htb]
\begin{center}
\includegraphics*[width=5cm]{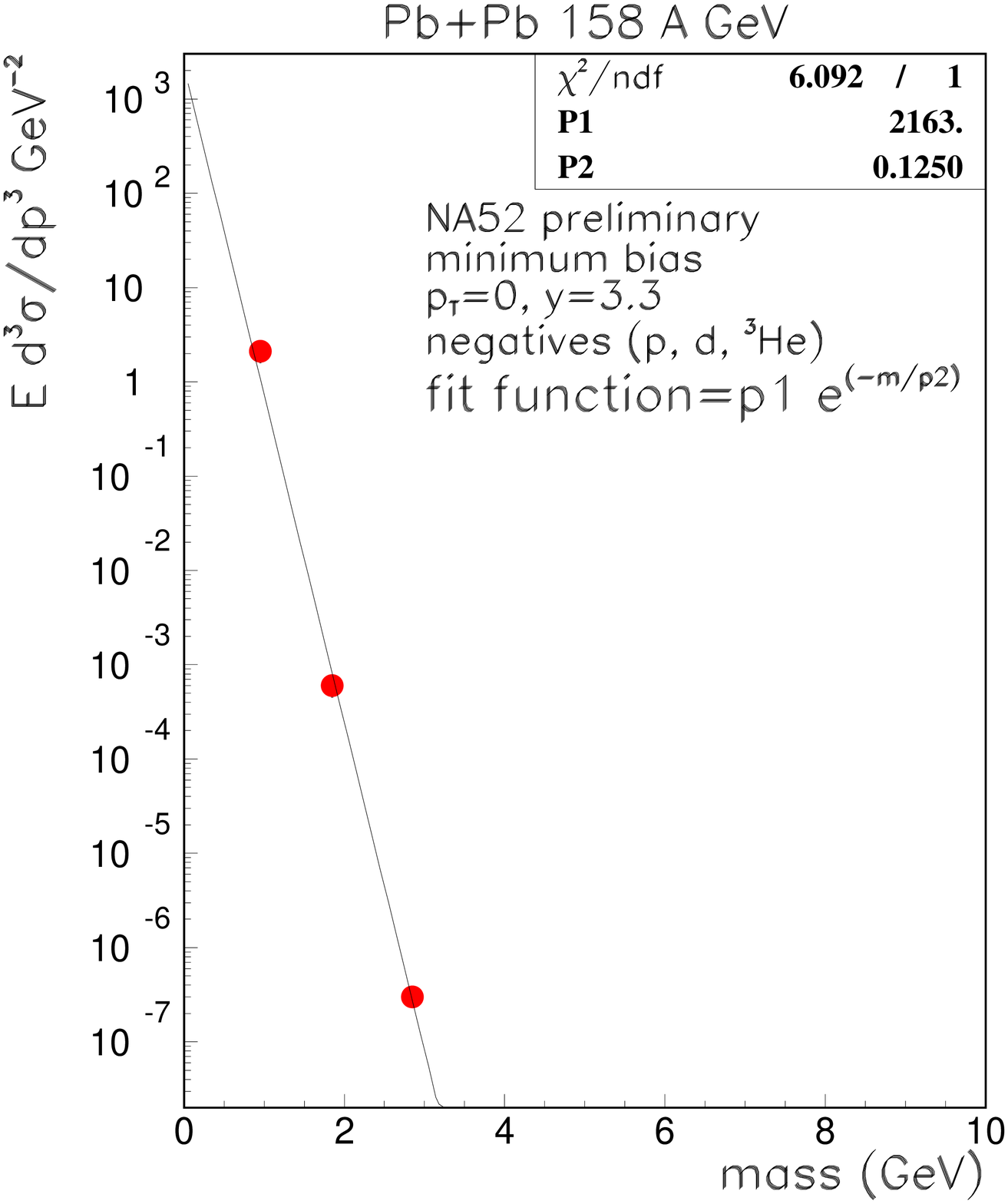}
\includegraphics*[width=5cm]{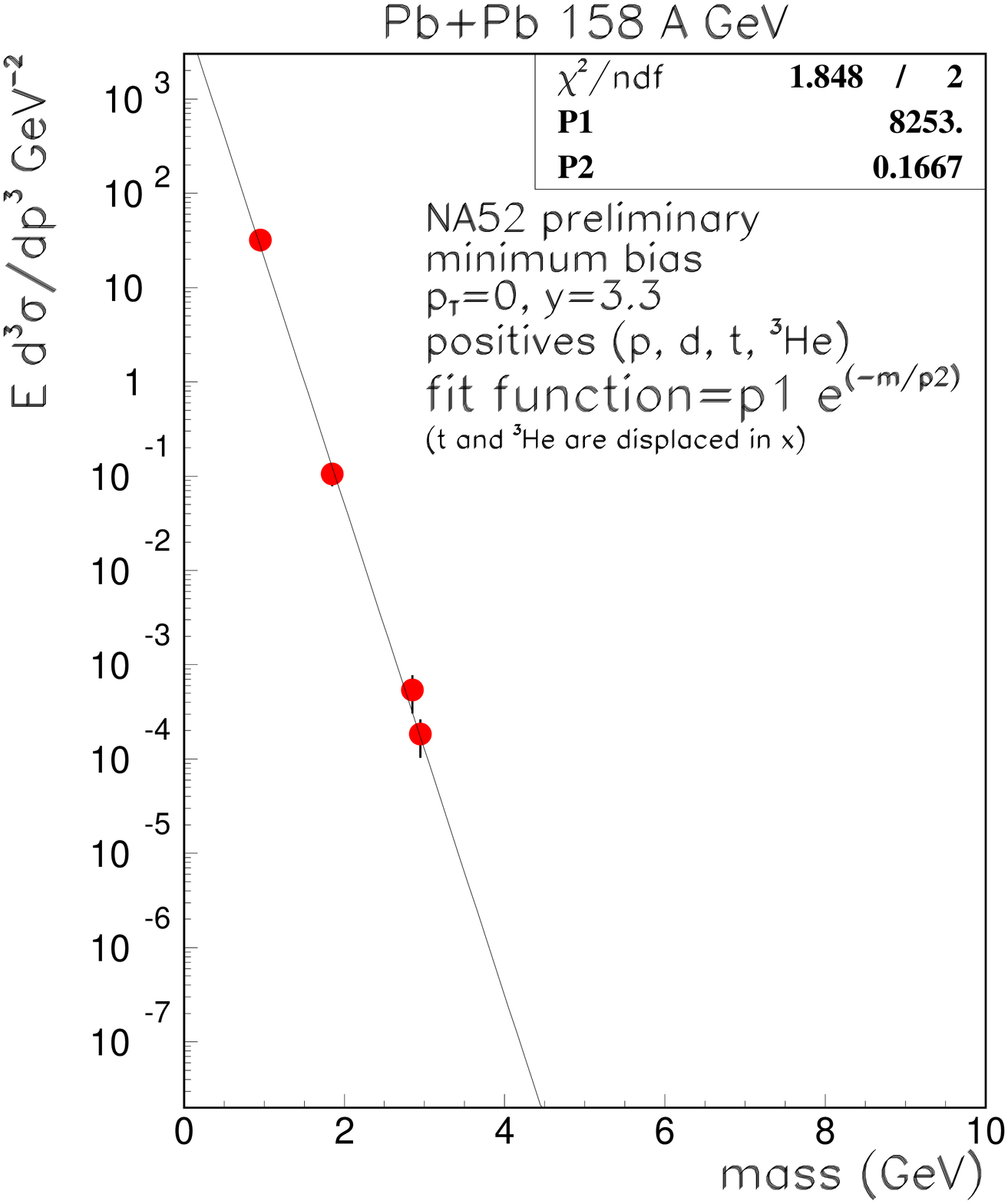}
\end{center}
\caption[]{
Proton and nuclei (left) and
 antiproton and antinuclei (right) cross sections as a function of mass in Pb+Pb
collisions at 158 A GeV taken with a minimum bias trigger, 
near zero $p_T$ and at y=3.3. 
The systematic errors have been quadratically added to the statistical ones.
The resulting  errorbars are of the size of the data points.
The straight lines represent a fit through the data points
using the function $f= p1 e^{ (-m/T) }$.
}
\label{thermal}
\end{figure}

\noindent
In the comparison of the antiparticle to particle yields to the simple
 coalescence model prediction, as shown in figure \ref{coal}, 
we assume that the positive and negative particles are produced through
the same production mechanisms.
We then examine whether this assumption is justified, by comparing
the so called coalescence factors $B_A$
for matter and antimatter.
This factor is defined as 
\[
B_A = \frac{Y_A}{(Y_{\mathrm{proton}})^A}
\]
with $A$ the atomic number and $Y$ the invariant differential yields.
\noindent
Figure \ref{compil} 
shows the coalescence scaling factors for central and minimum bias
events from the NA52 experiment together with other experimental data
from the Bevalac \cite{bev1,bev2}, BNL \cite{bnl1,bnl2,bnl3,bnl4} and the 
SPS \cite{cern1} experiments.
The NA52 coalescence scaling factors
 from $d/p^2$ and $\overline{d}/ \overline{p}^2$
 have been corrected for $\Lambda$ and $\overline{\Lambda}$ decays.
\\

\noindent
Several observations can be made from figures \ref{compil}:
\\
Firstly, the coalescence factor is similar for matter and antimatter
in accordance with the assumption that nuclei and antinuclei 
 are produced through the same  mechanism.
It supports the description of coalescence production.
Furthermore, 
the NA52 coalescence factors agree with data from the NA44 and NA49 experiments
at the same trigger conditions, $p_T$ and after the same weak decay corrections.
The reduction of the coalescence factor with increasing beam energy
for heavy ion collisions, seen in  figures  \ref{compil}
can be explained by an enhancement in the particle source volume
due to expansion.
\\

\noindent
Secondly, 
the coalescence factors decrease with increasing centrality
of the collisions.
This  can be  due to an increase in the particle
source volume, as more nucleons from the colliding nuclei
are involved in the interaction.
This tendency is also seen in figure \ref{r1} (left), where
 the coalescence factors $B_2$ and $\overline{B_2}$
as a function of the number of participant nucleons N 
are shown \cite{padova1}.
\\

\noindent
Under this assumption
one can extract the volume of the particle source using a coalescence model
\cite{llope}.
The radii infered from the $d/p^2$ and $\overline{d} / \overline{p}^2$
ratios  increase with the number of 
participant nucleons N as $\sim$ $N^{1/3}$, as shown in figure
\ref{r1} (right) \cite{padova1}.
They are compared to the source radii obtained from $\pi \pi$
interferometry in Fig. \ref{t_vs_n} (left).
The data are consistent with a transverse mass dependence 
of the source radii due to transverse expansion of the particle
source \cite{scheibl}.
\\

\noindent
Assuming chemical and thermal equilibrium for baryons and antibaryons
 we can infer from the measured particle yields 
information about the temperature $T$ and baryochemical potential $\mu_{\mathrm b}$ at the chemical freeze out.
We assume that the particles fill the phase space according to a Boltzmann distribution.
Thus we can write the invariant differential cross section for particles with a chemical potential $\mu$, a spin $S$ and an energy $E$ as:

\begin{equation}
 E \frac {d^3\sigma}{dp^3} = E\cdot(2S+1)\cdot 
\sigma_{PbPb}\cdot\frac{V}{(2\pi)^3} \exp{ (-\frac{E-\mu}{T} ) }
\end{equation}

\noindent
Here $V$ and $T$ are the volume and temperature of the source, respectively.

\noindent
In addition we 
assume that the volume and the temperature of the source is the 
same for all considered particles and that the chemical potentials of 
baryons ($\mu_{\mathrm b}$), 
antibaryons ($\mu _{\overline{ \mathrm b} }$) and nuclei ($\mu _A$) are related as follows:

\begin{equation}
\mu_{\overline{ \mathrm b} } = - \mu_{\mathrm b} \hspace*{1cm} ; \hspace*{1cm} \mu_A = A \cdot \mu_{\mathrm b}.
\end{equation}

\noindent
From the cross section ratios of $d/p$, $\overline{d}/ \overline{p}$
 and $\overline{p}/p$ 
 we obtain $\mu_{\mathrm b}$ and $T$. 
The latter parameters are shown
in figure \ref{t_vs_n} (right), as a function of the mean number
of  participant nucleons in the collision \cite{padova1}.
We find that the temperature extracted using the ratios
 $d/p $, $\overline{d} / \overline{p}$ and $ \overline{p} / p$, is
$\sim$ 125 MeV for the most central  collisions.
This temperature is the same as the one characterizing the thermal freeze-out
of hadrons \cite{heinz_qm99}.
This finding supports the assumption that $d$ and $\overline{d}$
(and maybe $\overline{p}$ too, as expected in \cite{shuryak})
freeze-out chemically, at the time of the thermal freeze-out
of hadrons. 
It suggests that in pairs produced nuclei and antinuclei
break up due to collisions in the phase between
chemical and thermal freeze-out.
Only the ones which form at a time close to thermal freeze-out of the other
hadrons, have a chance to survive.
\\

\noindent
In Fig. \ref{nagle1} the production cross sections for baryons
as a function of their mass are shown for central and forward rapidities
y=3.3 and y=5.7 respectively.
The data points are fit with a function $f = p1/p2^{(m-1)}$
\cite{nagle}
from which the penalty in the production cross sections
for higher mass nuclei can be extracted.
The penalty  factors f for positively charged baryons
at midrapidity comes out to be f=378 per baryon, and at forward rapidity 
f=24.
A similar fit  to the antibaryon cross sections
is shown in fig. \ref{nagle2}.
The obtained penalty  factor  is f=2968.
The penalty  factor for baryons is smaller than for antibaryons. 
This is due to the
feeding of baryons from the fragmentation into the midrapidity region
which is expecially effective at low $p_T$.
The penalty factors found in our experiment are different
from the ones extracted from the AGS experiment
E864 \cite{nagle}.
This is mainly due to the different
energies of the incident ions 
and the different $p_T$ acceptance of the detectors.
\\

\noindent
In figure \ref{thermal} we compare the nuclei and antinuclei data
with the expectation of a thermal model.
The inverse slope parameter extracted from 
a fit with the function $f=p1 e^{(-m/T)}$
through the antibaryon production cross sections
  at midrapidity turns out to give T=125 MeV 
(however the $\chi^2/DOF$ is not so good).
The baryons at midrapidity however have a temperature of
 T=167 MeV (with a good $\chi^2/DOF$).
This difference is due to feeding 
 from target and projectile fragments
as described above.
The temperature T=125 MeV found from the inverse slope parameter
turns out to be consistent with the chemical freeze-out temperature
obtained from the $d/p$, $\overline{d}/ \overline{p}$
 and $\overline{p} / p$ ratios
and very close to the thermal freeze-out temperature T=120 MeV
of other hadrons.

\section{Summary}

NA52 investigates particle and antiparticle
production in Pb+Pb collisions at 158 A GeV.
Two $\overline{ ^3He}$ and $10^3$ $\overline{d}$ were observed.
The decrease of the $\overline{p}/p$ and the $\overline{d}/d$ ratios
with increasing centrality of the collision, 
suggests annihilation and breakup processes.
The coalescence picture describes the Pb+Pb data well.
The coalescence factors $B_A$ decrease with increasing centrality of
the collision due to an increasing source size.
The extracted radius from the $d/p^2$ ratio
is in agreement with pion interferometry results.
The radii decrease with increasing $m_T$ as expected for
a transversally expanding source.
\\

\noindent
Using a thermal and a coalescence model we extract
a thermal and a chemical freeze-out temperature for 
baryons and antibaryons of T=125 MeV.
This temperature is very close to the thermal freeze-out
temperature of hadrons at T=120 MeV.
The results support
the picture that the surviving nuclei and antinuclei
are dominantly formed via coalescence at a time very close to
the thermal freeze-out of hadrons,
while those forming earlier
are mostly destroyed due to annihilation and breakup
reactions in the hadron dense environment.
\\

\section*{Acknowledgements}
We thank the Schweizerischer Nationalfonds for their support.

\vfill\eject
\end{document}